\documentclass[letterpaper,onecolumn,12pt,accepted=2024-10-07]{quantumarticle}
\pdfoutput=1 
\usepackage[utf8]{inputenc}
\usepackage[english]{babel}
\usepackage[T1]{fontenc}
\usepackage{xkeyval}
\usepackage{etoolbox}
\usepackage{geometry}
\usepackage{fancyhdr}
\usepackage{tikz}
\usepackage{ltxcmds}

\usepackage{amssymb}
\usepackage{amsmath}
\usepackage[space]{cite}
\usepackage{bbm}
\usepackage{hyperref}
\usepackage{graphicx}
\usepackage{adjustbox}
\usepackage{caption}
\usepackage{subcaption}
\usepackage{xcolor}
\usepackage[normalem]{ulem}

\usepackage{cases}

\numberwithin{equation}{section}

\newcommand{\non}{\nonumber}
\newcommand{\id}{\mathbb{I}}

\newcommand{\RN}[1]{%
  \textup{\uppercase\expandafter{\romannumeral#1}}%
}

\newcommand{\cut}[1]{\ifmmode\text{\textcolor{red}{\sout{\ensuremath{#1}}}}\else\textcolor{red}{\sout{#1}}\fi}

\usepackage{quantikz}

\begin{document}

\title{Deterministic Bethe state preparation}

\author{David Raveh}
\affiliation{Department of Physics, PO Box 248046, University of Miami, Coral Gables, FL 33124 USA}
\email{david.raveh@rutgers.edu}

\author{Rafael I. Nepomechie}
\affiliation{Department of Physics, PO Box 248046, University of Miami, Coral Gables, FL 33124 USA}
\email{nepomechie@miami.edu}

\maketitle

\begin{abstract}
We present an explicit quantum circuit that prepares an arbitrary
$U(1)$-eigenstate on a quantum computer, including the exact
eigenstates of the spin-1/2 XXZ quantum spin chain with either open or
closed boundary conditions.  The algorithm is deterministic, does not
require ancillary qubits, and does not require QR decompositions.  The
circuit prepares such an $L$-qubit state with $M$ down-spins using
$\binom{L}{M}-1$ multi-controlled rotation gates and $2M(L-M)$ CNOT-gates.
\end{abstract}

\setcounter{footnote}{0}

\section{Introduction}\label{sec:intro}

It is widely believed that the simulation of quantum physics is one of
the most promising applications of quantum computers, see e.g.
\cite{Preskill:2021apy, Aaronson:2022fwt}.  Among the potential target quantum systems,
one-dimensional quantum spin chains are attractive candidates.
Indeed, these are many-body quantum systems that appear in a myriad of
contexts in fields such as physics (condensed matter
\cite{Giamarchi2004}, statistical mechanics \cite{Baxter1982,
Mallick:2015}, high-energy theory \cite{Beisert:2010jr}), chemistry
\cite{Tilly:2021jem} and computer science \cite{Gharibian:2019}.  A
subset of those models are quantum integrable, and thus their exact
energy eigenstates (``Bethe states'') and eigenvalues can be expressed
in terms of solutions (``Bethe roots'') of so-called Bethe equations.
These results can be derived by either the coordinate
\cite{Bethe:1931hc, Gaudin:1971zza, Gaudin:1983, Alcaraz:1987uk} or
algebraic \cite{Faddeev:1981ft, Korepin:1993, Sklyanin:1988yz} Bethe
ansatz.  Given the Bethe roots (say, of the ground state), it would be
desirable to prepare the corresponding Bethe state on a quantum
computer \cite{Nielsen:2019}, which would then allow the computation
of correlation functions in this state, see e.g. \cite{Somma:2002,
Wecker:2015}.

After \cite{Nepomechie:2021bethe}, an algorithm for preparing Bethe
states of the spin-1/2 closed XXZ chain based on the coordinate Bethe
ansatz was formulated \cite{VanDyke:2021kvq}.  This algorithm, which
was later generalized to the open chain \cite{VanDyke:2021nuz}, is
restricted to real Bethe roots, requires ancillary qubits, and is
probabilistic.  Moreover, the success probability was shown to tend
super-exponentially to zero with the number of down-spins \cite{Li:2022czv}.  A different approach, based instead on the algebraic Bethe ansatz, was subsequently developed for the closed XXZ chain
\cite{Sopena:2022ntq}.  The latter algorithm is not
limited to real Bethe roots, and is deterministic; however, it is not explicit, as it requires performing QR decompositions of matrices whose size scales exponentially with the number of down-spins. Analytic formulae for the unitaries of \cite{Sopena:2022ntq} were proposed in \cite{Ruiz:2023rew}. 

We present here a new algorithm for preparing arbitrary $L$-qubit
$U(1)$-eigenstates, including exact Bethe states of the
spin-1/2 XXZ chain with either open or closed boundary conditions.
Like the algorithms \cite{VanDyke:2021kvq, VanDyke:2021nuz}, it is
based on the coordinate Bethe ansatz; however, it is not limited to real
Bethe roots, does not require ancillary qubits, and it is
deterministic.  Moreover, unlike \cite{Sopena:2022ntq},
our algorithm is explicit, and does not make use of QR
decompositions.  For a state with $M$ down-spins in a chain of length
$L$, the circuit uses only $L$ qubits, and has size (gate count) and depth $\mathcal{O}\left(\binom{L}{M}\right)$. The algorithm makes use of certain complex numbers $f(w)$, which are expensive to compute classically for large $M$. Our algorithm is inspired by
an efficient recursive algorithm\cite{Bartschi2019,
Nepomechie:2023lge, Raveh:2023iyy, Nepomechie:2024fhm} for preparing
Dicke states \cite{Dicke:1954zz}.  Code in Qiskit \cite{Qiskit}
implementing our algorithm is available as Supplementary Material.

The remainder of this paper is structured as follows. In Section \ref{sec:recursion}, starting from an expression for the Bethe state that we wish to prepare, we define the states \eqref{psidef}, and observe that they satisfy
an elementary recursion \eqref{psirec}. In Section \ref{sec:algorthim}, we use this recursion to derive a corresponding recursion for a unitary operator $U_L$ that prepares the Bethe state from a simple product state \eqref{Bethestate},
in terms of certain simpler unitary operators $W_m$, see \eqref{ULresult}.  The latter operators are constructed in terms of yet simpler unitary operators $\RN{1}_{m,l}$, see \eqref{Wmresult}. A brief discussion of these results is presented in Section \ref{sec:discuss}. In Appendix \ref{sec:Bethe},
the coordinate Bethe ansatz for the closed and open spin-1/2 XXZ chains is  summarized. Finally, complete circuit diagrams for examples with $(L,M)=(4,2)$ and $(L,M)=(6,3)$ are presented in Appendix \ref{sec:examples}.

\section{Recursion}\label{sec:recursion}

We consider the eigenstates of any Hamiltonian with $U(1)$ symmetry.
An important example are Bethe states, which are eigenstates
of the XXZ spin-1/2 quantum spin chain of length $L$ with either open
or closed (periodic) boundary conditions.  In the coordinate Bethe
ansatz approach \cite{Bethe:1931hc, Gaudin:1971zza, Gaudin:1983, Alcaraz:1987uk}, a normalized Bethe state with $0<M<L$ down-spins is
expressed as
\begin{equation}
    |B^L_M\rangle=\frac{1}{F(\{\})}\sum_{w\in P(L,M)}f(w)\,|w\rangle\,,
    \label{BLM}
\end{equation}
where the sum is over the set $P(L,M)$ of all permutations $w$ with $L-M$ zeros (up-spins) and $M$ ones (down-spins), which we think of as $\binom{L}{M}$ strings of length $L$.
Moreover, the coefficients $f(w)$ are complex numbers that depend on the corresponding $M$ Bethe roots, which we assume are known, see Appendix \ref{sec:Bethe}. We emphasize that these coefficients are computed classically, with complexity that scales exponentially 
with $M$. Here $F(\{\})$ denotes the normalization
\begin{equation}
\label{F({})}
    F(\{\})=\sqrt{\sum_{w\in P(L,M)}|f(w)|^2}\,,
\end{equation}
and the motivation for this notation shall soon become clear. For example,
\begin{align}
|B^4_2\rangle=\frac{1}{F(\{\})}\big(
&f(0011)\,|0011\rangle+
f(0101)\,|0101\rangle+
f(1001)\,|1001\rangle\nonumber\\+
&f(0110)\,|0110\rangle+
f(1010)\,|1010\rangle+
f(1100)\,|1100\rangle\big)\,,
\label{B42}
\end{align}
and $|0011\rangle=|0\rangle\otimes|0\rangle\otimes|1\rangle\otimes|1\rangle$, etc., with computational basis states $|0\rangle=\begin{pmatrix}1\\0\end{pmatrix}$ and $|1\rangle=\begin{pmatrix}0\\1\end{pmatrix}$. The Bethe states are not the only examples of states that take the form \eqref{BLM}; in fact, the eigenstates of any Hamiltonian with $U(1)$ symmetry take this form.\footnote{Indeed, since the states $|B^L_M\rangle$ form an orthonormal basis, any state $|\phi\rangle$ can be expressed as
$|\phi\,\rangle=\sum_{M=0}^L C_M\, |B^L_M\rangle$. Hence, if 
$|\phi\rangle$ is an eigenstate of $S^z$  with eigenvalue $\alpha$, 
then $\alpha\,|\phi\rangle=S^z|\phi\rangle=\sum_{M=0}^L\, 
(L/2-M)\,C_M\, |B^L_M\rangle$. This implies $\alpha 
\,C_M=(L/2-M)\,C_M$, so either $\alpha=L/2-M$ or $C_M=0$. In particular, $C_M$ is non-zero for exactly one value of $M$, so $|\phi\rangle = |B^L_M\rangle$.} Although we have in mind values $f(w)$ dictated by Bethe ansatz, and we shall refer to $|B^L_M\rangle$ as Bethe states, our algorithm is actually more general, as it prepares the state $|B^L_M\rangle$ for \emph{arbitrary} complex numbers $f(w)$. In particular, for the case $f(w)=1$, $|B^L_M\rangle$ reduces to the Dicke state $|D^L_M\rangle$, whose preparation
was shown to have an efficient recursive algorithm \cite{Bartschi2019}.

Inspired by \cite{Bartschi2019, Nepomechie:2023lge, Raveh:2023iyy, Nepomechie:2024fhm}, we shall prepare the Bethe state $|B^L_M\rangle$ recursively. To do so, for a string $b$ of zeros and ones, let us define the state
\begin{equation}
|\psi(b)\rangle=\frac{1}{F(b)}
\sum_{\substack{w\in P(L,M)\\w=ab}}
f(w)\,|w\rangle\,,
\label{psidef}
\end{equation}
with normalization
\begin{equation}
F(b)=
\begin{cases}
\sqrt{
\sum\limits_{\substack{w\in P(L,M)\\w=ab}}
|f(w)|^2}\,,& \text{multiple $a$ satisfy }w=ab \in P(L,M)\,, \\
f(w)\,,& \text{unique $a$ satisfies }w=ab \in P(L,M) \,.
\end{cases}
\end{equation}
That is, we are restricting the sum in \eqref{psidef} to permutations $w$ of the form $w=ab$, where $ab$ denotes the {\em concatenation} of the strings $a$ and $b$, so $|w\rangle=|a\rangle\otimes|b\rangle$. The state \eqref{psidef} is thus the ``$b$-tail'' of the Bethe state \eqref{BLM}. In other words, the sum in \eqref{psidef} is effectively over $a$, while keeping $b$ fixed.
Note that when $b=\{\}$ is the empty string, $F(b)$ reduces to the normalization given by \eqref{F({})}.  
For example, for the state $|B^4_2\rangle$ \eqref{B42} with $b=0$, we have
\begin{equation}
    |\psi(0)\rangle= \frac{1}{F(0)}\left(
f(0110)\,|0110\rangle+
f(1010)\,|1010\rangle+
f(1100)\,|1100\rangle\right) \,.
\label{ex42}
\end{equation}

We observe that the states $|\psi(b)\rangle$ satisfy the elementary 
but important recursion
\begin{equation}
\label{psirec}
|\psi(b)\rangle=G(0b)\,|\psi(0b)\rangle+G(1b)\,|\psi(1b)\rangle\,,
\end{equation}
with
\begin{equation}
\label{Gib}
G(ib)=\frac{F(ib)}{F(b)}\,, \qquad i = 0,1 \,,
\end{equation}
where $ib$ (with $i=0,1$) is the concatenation of the strings $i$ and $b$. For the example \eqref{ex42},
we have 
\begin{align}
   |\psi(0)\rangle&=G(00)\,|\psi(00)\rangle+G(10)\,|\psi(10)\rangle\non
   \\&=\frac{G(00)}{F(00)}\,f(1100)\,|1100\rangle +
   \frac{G(10)}{F(10)}\left(
f(0110)\,|0110\rangle+
f(1010)\,|1010\rangle\right) \,.
\end{align}
It is clear that $|B^L_M\rangle=|\psi(\{\})\rangle$ corresponding to the empty string $b=\{\}$, so preparing the state $|\psi(b)\rangle$ recursively over the length of the string $b$ amounts to a recursive preparation of the state $|B^L_M\rangle$.

\section{Algorithm}\label{sec:algorthim}

Suppose for $m\leq L$ the unitary operator $U_m$ independent of $l$ and $b$ satisfies
\footnote{
We occasionally put a subscript on a ket, such as $|\psi(b)\rangle_L$\,, to clarify that it is an $L$-qubit ket.}
\begin{equation}
U_m\,|0\rangle^{\otimes (m-l)}\,|1\rangle^{\otimes l}\,|b\rangle_{L-m}=|\psi(b)\rangle_L
\label{Um=psi}
\end{equation}
for all $0<l<m$ and all strings $b$ of length $L-m$ with $M-l$ ones, i.e. for all $b\in P(L-m,M-l)$. This implies the restriction $0\leq M-l\leq L-m$, so that 
\begin{equation}
    \max(M+m-L,1)\leq l\leq \min(m-1,M) \,.
    \label{lbounds}
\end{equation}
The reason we exclude the cases $l=m$ or $l=0$ is that in these cases, $w=ab\in P(L,M)$ is unique, and hence $U_m=\id^{\otimes L}$. This implies that for the case $m=1$, where $l=0$ or $l=m=1$, we must have $U_1=\id^{\otimes L}$. For the case $m=L$, we have $b=\{\}$, hence \eqref{Um=psi} implies that
\begin{equation}
|B^L_M\rangle=U_L\,\,|0\rangle^{\otimes (L-M)}\,|1\rangle^{\otimes M}\,.
\label{Bethestate}
\end{equation}

It follows from the recursion \eqref{psirec} that $U_m$ has the recursive property
\begin{align}
U_m\,|0\rangle^{\otimes (m-l)}\,|1\rangle^{\otimes l}\,|b\rangle_{L-m}&=
G(0b)\,U_{m-1}\,|0\rangle^{\otimes (m-l-1)}\,|1\rangle^{\otimes l}\,|0b\rangle_{L-m+1}\nonumber\\&
+
G(1b)\,U_{m-1}\,|0\rangle^{\otimes (m-l)}\,|1\rangle^{\otimes (l-1)}\,|1b\rangle_{L-m+1}\,.
\end{align}
This motivates finding a unitary operator $W_m$ independent of $l$ and $b$ that satisfies 
\begin{align}
W_m\,|0\rangle^{\otimes (m-l)}\,|1\rangle^{\otimes l}\,|b\rangle_{L-m}&=
G(0b)\,|0\rangle^{\otimes (m-l-1)}\,|1\rangle\,|1\rangle^{\otimes (l-1)}\,|0\rangle\,| b\rangle_{L-m}\nonumber\\&
+
G(1b)\,|0\rangle^{\otimes (m-l-1)}\,|0\rangle\,|1\rangle^{\otimes (l-1)}\,|1\rangle\,| b\rangle_{L-m}\,
\label{Waction}
\end{align}
for all $\max(M+m-L,1)\leq l\leq \min(m-1,M)$ and $b\in P(L-m,M-l)$. In terms of this operator, $U_m$ satisfies the recursion
\begin{equation}
U_m=U_{m-1}W_m
\end{equation}
which, using $U_1=\id^{\otimes L}$, can be telescoped to 
\begin{equation}
U_L=  \overset{\curvearrowright}{\prod_{m=2}^{L}} 
W_m\,,
\label{ULresult}
\end{equation}
with the product going from left to right with increasing $m$. 

It thus remains to construct a circuit $W_m$ that performs \eqref{Waction} for all $l$ and $b$. This can be achieved by expressing $W_m$ as an ordered product of simpler operators $\RN{1}_{m,l}$ independent of $b$ that perform the job of $W_m$ \eqref{Waction} for specific values $l$, in such a way that operators with different $l$-values do not interfere with each other. That is, $\RN{1}_{m,l}$
has the following properties:
\begin{align}
    \RN{1}_{m,l'}\,|0\rangle^{\otimes (m-l)}\,|1\rangle^{\otimes l}\,|b\rangle_{L-m} &= \begin{cases}
        \quad\,\,\,\,\, |0\rangle^{\otimes (m-l)}\,|1\rangle^{\otimes l}\,|b\rangle_{L-m} & l' < l\\
        W_m \,|0\rangle^{\otimes (m-l)}\,|1\rangle^{\otimes l}\,|b\rangle_{L-m} & l'=l
    \end{cases} \,, \label{Iprop1} \\
   \RN{1}_{m,l'} \left( \RN{1}_{m,l}\,|0\rangle^{\otimes (m-l)}\,
   |1\rangle^{\otimes l}\, |b\rangle_{L-m}  \right) &= 
    \RN{1}_{m,l}\,|0\rangle^{\otimes (m-l)}\,|1\rangle^{\otimes l}\,|b\rangle_{L-m} \qquad l' > l \,, \label{Iprop2}
\end{align}
for all $b$, where $W_m\, |0\rangle^{\otimes (m-l)}\,|1\rangle^{\otimes l}\,|b\rangle_{L-m}$ in \eqref{Iprop1} is given by \eqref{Waction}. 
Indeed, it readily follows from these properties that $\RN{1}_{m,l}$ performs the mapping \eqref{Waction} for fixed $l$, and that
\begin{equation}
    \left( \overset{\curvearrowleft}{\prod_{l'=1}^{m-1}}\RN{1}_{m,l'} \right)
    \,|0\rangle^{\otimes (m-l)}\,|1\rangle^{\otimes l}\,|b\rangle_{L-m} = W_m \,|0\rangle^{\otimes (m-l)}\,|1\rangle^{\otimes l}\,|b\rangle_{L-m} 
\end{equation}
for all $l$.  In view of the bounds \eqref{lbounds}, we conclude that
the operator $W_m$, independent of $l$, can be written as the
following ordered product of $\RN{1}$-operators:
\begin{equation}
W_m=
\overset{\curvearrowleft}{\prod\limits_{l=\max(M+m-L,1)}^{\min(m-1,M)}}
\,\RN{1}_{m,l} \,,
\label{Wmresult}
\end{equation}
with the product going from right to left with increasing $l$.

\begin{figure}[htb]
	\centering
	\begin{subfigure}{0.5\textwidth}
      \centering
\begin{adjustbox}{width=0.5\textwidth, raise=5.8em}
\begin{quantikz}
\lstick{$0$}&\qw &\gate[3,disable auto 
height]{\substack{ {\textstyle\bullet} \\ \\
{\textstyle\bullet}\\ \\ {\textstyle\bullet} }} &\qw &\qw \\
\vdots & \qw & \qw & \qw & \qw \\
\lstick{$L-m-1$}&\qw&\vqw{1}\qw&\qw&\qw\\
\lstick{$L-m$} & \ctrl{1}  &  \gate{\mathcal{U}(m,l)} \vqw{1} & 
\ctrl{1}  & \qw \\
\lstick{$L-m+1$} & \targ{} &  \ctrl{-1} & 
\targ{}   & \qw \\
\vdots &&&&\\
\lstick{$L-1$}& \qw & \qw & \qw & \qw \\
\end{quantikz}
\end{adjustbox}
\caption{$\RN{1}_{m,l}$ with $l=1$}
\label{fig:I1}
    \end{subfigure}%
    \begin{subfigure}{0.5\textwidth}
        \centering
\begin{adjustbox}{width=0.5\textwidth}
\begin{quantikz}
\lstick{$0$}&\qw&\gate[3,disable auto 
height]{\substack{{\textstyle\bullet}\\ \\
{\textstyle\bullet} \\ \\ {\textstyle\bullet} }} &\qw&\qw\\
\vdots & \qw & \qw & \qw & \qw \\
\lstick{$L-m-1$}&\qw&\vqw{1}\qw&\qw&\qw\\
\lstick{$L-m$} & \ctrl{1}  &  \gate{\mathcal{U}(m,l)} \vqw{1} & \ctrl{1}  & \qw \\
\vdots &&&&\\
\lstick{$L-m+l-1$} & \qw \vqw{-1} & \ctrl{-1} \qw & \qw \vqw{-1} & \qw \\
\lstick{$L-m+l$} & \targ{} \vqw{-1} &  \ctrl{-1}  & \targ{} \vqw{-1}   & \qw \\
\vdots &&&&\\
\lstick{$L-1$}& \qw & \qw & \qw & \qw \\
\end{quantikz}
\end{adjustbox}
\caption{$\RN{1}_{m,l}$ with $l>1$}
\label{fig:Il}
	 \end{subfigure}	
\caption{Circuit diagrams for $\RN{1}_{m,l}$, with $\mathcal{U}(m,l)$ defined as a product of $u$-gates \eqref{U(m,l)}. Additional controls on wires $0$ to $L-m-1$ (represented by a rectangle with 3 bullets)
are placed on each $u(m,l,b)$ at the locations where $b$ contains a $1$. The number of $1$'s in $b$ is $M-l$.}
\label{fig:Iops}
\end{figure}
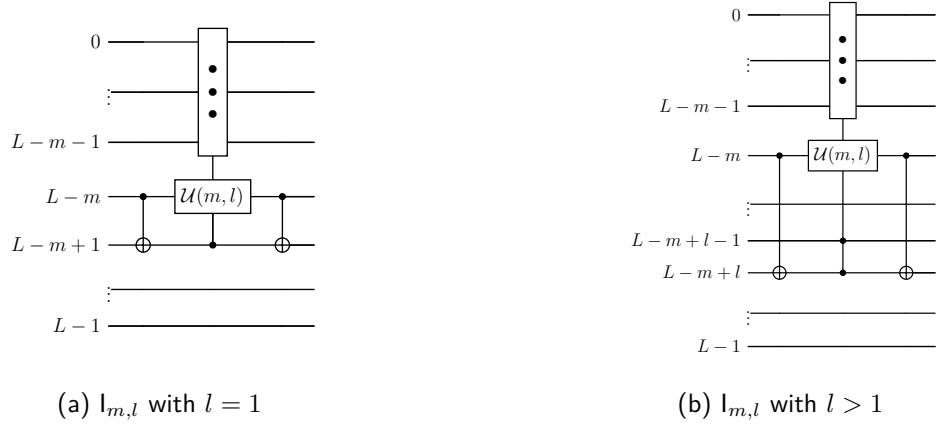

The circuit diagram for an operator $\RN{1}_{m,l}$ with the properties \eqref{Iprop1} and \eqref{Iprop2} is displayed in Fig. \ref{fig:Iops}.
Similarly to the Dicke circuit \cite{Bartschi2019,
Nepomechie:2023lge, Raveh:2023iyy, Nepomechie:2024fhm}, the pair of CNOT gates together with the multi-controlled $\mathcal{U}(m,l)$-gate perform the rotation in the subspace $|10\rangle$ and $|01\rangle$ defined by \eqref{Waction}. $\mathcal{U}(m,l)$ denotes the product
\begin{equation}
\mathcal{U}(m,l)=\prod_{b\in P(L-m,M-l)} u(m,l,b)\,
\label{U(m,l)}
\end{equation}
with the product over all $b\in P(L-m,M-l)$ in arbitrary 
order.\footnote{If $f(w)=0$ for some subset of permutations $w$, it is possible that $F(b)=0$ for some $b$, so that $G(ib)$ \eqref{Gib} is singular; in these cases, the gates $u(m,l,b)$ should simply be omitted. If the product in \eqref{U(m,l)} is empty, the operator $\RN{1}_{m,l}$ can be omitted.} Here $u(m,l,b)$ is the $u$-gate
\begin{equation}
 u(m,l,b) 
 = \begin{pmatrix}
       \cos(\frac{\theta}{2}) & -e^{i \lambda}\sin(\frac{\theta}{2}) \\
       e^{i \phi}\sin(\frac{\theta}{2}) & e^{i (\phi+\lambda)} \cos(\frac{\theta}{2})
\end{pmatrix} 
\label{ugate}
\end{equation}
whose angles $\theta, \phi, \lambda$
depend on $b$ as follows:
\begin{equation}
\theta = 2 \arccos\left(|G(1b)|\right)\,, \qquad 
\lambda = \arg(G(0b))-\pi\,,\qquad
\phi = \arg(G(1b))-\lambda\,. 
\label{angles}
\end{equation}
Each $u$-gate acts on wire $L-m$, and has controls on the wires $L-m+l$ and $L-m+l-1$ (the latter of which is present if and only if $l>1$).\footnote{
We follow Qiskit conventions, where e.g. $|110\rangle$ corresponds to $|0\rangle$ on wire $0$ and $|1\rangle$ on wires $1$ and $2$.
} Additionally, because the angles depend on $b$, additional controls are placed on wires $0$ to $L-m-1$ at the locations where the string $b$ takes the value $1$, to differentiate between different $b$.\footnote{This method has been used before to generate states with many free parameters, see e.g. \cite{Mottonen:2004dis}. It suffices to put controls for the $1$'s in $b$, as opposed to also including negated controls for $0$'s. This is because each $\RN{1}_{m,l}$ operator conserves the number of $1$'s, and only acts on states of the form $|0\rangle^{\otimes (m-l)}\,|1\rangle^{\otimes l}\,|b\rangle_{L-m}$, which guarantees that $b$ contains exactly $M-l$ $1$'s.} For example, for $L=4,\,M=2$, $u(m=2,l=1,b=01)$ has controls on the wires $3$ and $0$, whereas $u(m=2,l=1,b=10)$ has controls on the wires $3$ and $1$. For the case in which $b$ in \eqref{U(m,l)} is unique, these additional controls can be omitted. Complete circuit diagrams for examples with $L=4,\,M=2$ and $L=6,\,M=3$ are displayed in Figs. \ref{fullcircuit4,2} and  \ref{fullcircuit6,3} in  Appendix \ref{sec:examples}. An implementation of this algorithm using Qiskit \cite{Qiskit} is available as Supplementary Material.

To summarize, the Bethe state $|B^L_M\rangle$ is given by
\eqref{Bethestate}, where $U_L$ is given by \eqref{ULresult} in terms
of $W_m$ \eqref{Wmresult}, where the operators $\RN{1}_{m,l}$ are
given by Fig.  \ref{fig:Iops}.  The $\mathcal{U}$-gate in Fig.
\ref{fig:Iops} defined in \eqref{U(m,l)} is a product of $u(m,l,b)$
gates \eqref{ugate}, with additional controls on the wires $0$ to
$L-m-1$ at the locations of the ones in the string $b$.

The number of CNOT-gates in $U_L$, as follows from Eqs. \eqref{ULresult} and \eqref{Wmresult}, is given by 
\begin{equation}
\sum_{m=2}^L \,\,\sum_{l=\max(M+m-L,1)}^{\min(m-1,M)} 2\,=2M(L-M)\,.
\label{cnotnum}
\end{equation}
The number of $u$-gates in $U_L$, considering \eqref{U(m,l)}, is given by
\begin{equation}
\sum_{m=2}^L \,\,\sum_{l=\max(M+m-L,1)}^{\min(m-1,M)}{L-m \choose M-l}={L \choose M}-1\,.
\label{ugatenum}
\end{equation}
We see that this circuit generates the state $|B^L_M\rangle$, which
has $\binom{L}{M}$ free parameters $f(w)$, $w\in P(L,M)$, with
precisely $\binom{L}{M}-1$ rotations.   The total circuit
size, including preparing the initial state $|0\rangle^{\otimes
(L-M)}|1\rangle^{\otimes M}$, is therefore
\begin{equation}
    M + 2M(L-M) + {L \choose M}-1 = \mathcal{O}\left(\binom{L}{M}\right) \,,
\end{equation}
and the depth is
$\mathcal{O}\left(\binom{L}{M}\right)$ as well. 
For the central case $M=L/2$, we note the asymptotic approximation (see e.g. \cite{Luke:1969}, p. 35)
\begin{equation}
\binom{L}{L/2} = \frac{2^L}{\sqrt{\pi L/2}}\left(1  + \mathcal{O}(1/L) \right) \,.
\end{equation}
Further, we see that the number
of CNOT-gates \eqref{cnotnum} and $u$-gates \eqref{ugatenum} are symmetric under $M\to L-M$, which
implies that there is no advantage to using the dual symmetry
$|B^L_M\rangle=X^{\otimes L}\,|B^L_{L-M}\rangle$. 

For the special case that all $f(w)=1$, corresponding to Dicke states, our circuit is similar to, but does not reduce to, the one in \cite{Bartschi2019, Nepomechie:2023lge, Raveh:2023iyy, Nepomechie:2024fhm}. This is because the rotation angles in  $u(m,l,b)$ are the same for all $b$, so only $M(L-M)$ total rotations are needed, and the additional controls on the wires $0$ to $L-m-1$ are unnecessary. 

\section{Discussion}\label{sec:discuss}

We have formulated an algorithm for exactly preparing on a quantum
computer an $L$-qubit $U(1)$-eigenstate consisting of an arbitrary
superposition of computational basis states with $M$ ones (down-spins) and $L-M$ zeros (up-spins) \eqref{BLM}.  The circuit is ideally suited for models solved by Bethe ansatz, which have available general
recipes for the values $f(w)$ corresponding to the models' eigenstates
for arbitrary values of $L$ and $M$ in terms of Bethe roots $\{ k_1,
..., k_M \}$. However, these coefficients $f(w)$, of which there are ${L \choose M}$ many, are each expressed as a sum over $M!$ terms for the closed chain \eqref{fXXZclosed}, and $2^M M!$ terms for the open chain \eqref{fXXZopen}, and are thus expensive to compute classically for large $M$. 
(For eigenstates of non-integrable models, the $f(w)$
values are generally not known.)  The circuit size is of the same
order as the number of free parameters $f(w)$, suggesting that the
circuit with generic values of $f(w)$
is optimal for given values of $L$ and $M$.  Nevertheless, for
$M=L/2$, the circuit size and depth scale exponentially with $L$. As noted in the Introduction, these states could be used to compute correlation functions \cite{Somma:2002, Wecker:2015}.

For the isotropic ($\Delta=1$) closed chain, the Bethe states are $su(2)$ highest-weight states \cite{Faddeev:1981ft}. It should also be possible to prepare the descendant (lower-weight) states using functions $f(w)$ obtained by taking appropriate limits of Bethe roots \cite{Faddeev:1981ft}. 
Although we have focused here on eigenstates of XXZ models, we expect 
that this algorithm will also be applicable to other $U(1)$-invariant spin-1/2 models, such as the Lipkin-Meshkov-Glick \cite{Lipkin:1964yk} model.

We have assumed here, as in \cite{VanDyke:2021kvq, VanDyke:2021nuz,
Li:2022czv, Sopena:2022ntq, Ruiz:2023rew}, that the Bethe roots for
the desired state are known.  Real Bethe roots can readily be
determined classically, even for values $L\,, M \sim 10^3$, by iteration
(see, e.g. \cite{Alcaraz:1987zr} and references therein).  However,
complex Bethe roots are generally difficult to find classically.
Following up on an idea suggested in \cite{Nepomechie:2021bethe}, it
may be possible to implement a hybrid classical-quantum approach (see
e.g. \cite{Tilly:2021jem} and references therein) for determining such
complex Bethe roots by using our Bethe states as trial states and the
unknown Bethe roots as variational parameters.\footnote{This approach has now indeed been implemented in \cite{Raveh:2024cfh}.}

We emphasize that our algorithm makes use of integrability only in so far as computing $f(w)$ \eqref{fXXZclosed},  \eqref{fXXZopen}, which is performed classically. It would be interesting to find a way to further exploit integrability. We also note that, for given values of $L$ and $M$, the circuit size and depth are independent of the values of $f(w)$. For example, for Dicke states (all $f(w)=1$), our circuit does not reduce to the efficient one in \cite{Bartschi2019}. It would be desirable to find a way to take advantage of simplifications of $f(w)$, which also arise in matrix product state (MPS) representations, the XX chain, etc. While MPS methods give an efficient way to prepare ground states of gapped models \cite{Schollwock:2011}, implementing these approaches for excited states is more challenging with tensor networks. Further, such methods require an ancillary space that grows with the system size, as well as numerical decompositions of unitaries into elementary gates; this typically makes such algorithms non-explicit.

It would be interesting to generalize the approach presented here to integrable $U(1)$-invariant spin chains of higher spin \cite{Crampe:2011} and of higher rank \cite{Sutherland:1975vr, Sutherland:1985}, which should generalize the circuits for corresponding Dicke states \cite{Nepomechie:2024fhm}, \cite{Nepomechie:2023lge} respectively.
It would also be interesting to try to extend this work to 
models without $U(1)$ symmetry, such as the XYZ spin chain \cite{Baxter1982, Gaudin:1983}. Indeed, a recursion analogous to \eqref{psirec} should still hold for such eigenstates; hence, it may still be possible to formulate an appropriate algorithm.

\section*{Acknowledgments} 

We thank Esperanza L\`opez, Sergei Lukyanov,
Eric Ragoucy, 
Roberto Ruiz, 
Germ\`an Sierra,
Alejandro Sopena and Michelle Wachs for helpful correspondence or discussions.
RN is supported in part by the National Science Foundation under grant PHY 2310594, and by a Cooper fellowship. 

\appendix

\section{Bethe ansatz}\label{sec:Bethe}

We briefly summarize here the coordinate Bethe ansatz for the spin-1/2 XXZ quantum spin chain of length $L$. 

\subsection{Closed periodic XXZ chain}

The Hamiltonian of the closed periodic spin-1/2 XXZ chain is given by
\begin{equation}
    \mathcal{H} = -\tfrac{1}{2}\sum_{n=1}^L 
    \left(\sigma^x_n \sigma^x_{n+1} + \sigma^y_n \sigma^y_{n+1} + \Delta \left( \sigma^z_n \sigma^z_{n+1} - \id \right) \right) \,, \qquad \vec{\sigma}_{L+1} = \vec{\sigma}_1 \,,
\label{Hclosed}
\end{equation}
where as usual $\sigma^x_n, \sigma^y_n, \sigma^z_n$ are Pauli matrices at site $n$, and $\Delta$ is the anisotropy parameter. This Hamiltonian has the $U(1)$ symmetry
\begin{equation}
    \left[  \mathcal{H} \,, S^z \right] = 0 \,, \qquad S^z = \sum_{n=1}^L \tfrac{1}{2}\sigma^z_n \,.
\end{equation}
For $\Delta=1$, the model reduces to the isotropic $su(2)$-invariant XXX spin chain. 
Given a solution $\{ k_1, \ldots, k_M \}$ of the Bethe equations (see e.g. \cite{Baxter1982, Gaudin:1983})
\begin{equation}
    e^{i k_j L} =  \prod_{l=1; l\ne j}^M \left(- \frac{s(k_l, k_j)}{s(k_j, k_l)} \right) \,, \qquad j = 1, \ldots, M \,,
    \label{BEclosed}
\end{equation}
where
\begin{equation}
s(k, k') = 1 - 2 \Delta e^{i k'} + e^{i(k+k')} \,,
\label{s(k,k')}
\end{equation}
one can construct a corresponding exact simultaneous eigenvector 
$|B^L_M\rangle$ \eqref{BLM} of $\mathcal{H}$ and $S^z$, with 
\begin{equation}
    f(w) = \sum_{\sigma\in S_M} \varepsilon(\sigma)\, A(k_{\sigma(1)}, \cdots, k_{\sigma(M)})\, 
    e^{i \sum_{j=1}^M k_{\sigma(j)} x_j} \,,
\label{fXXZclosed}
\end{equation} 
where the sum is over the set $S_M$ of all permutations $\sigma$ of $\{1,\dots,M\}$, and $\varepsilon(\sigma)=\pm 1$ denotes the sign of the permutation $\sigma$. Moreover,
\begin{equation}
A(k_1, \ldots, k_M) = \prod_{1\le j<l\le M}  s(k_l, k_j) \,, 
\label{Acoeffclosed}
\end{equation} 
and the $x_j \in \{1, \ldots,L\} $ in \eqref{fXXZclosed} are the positions of the 1's in the argument $w$ of $f(w)$. For example, if $w=0101$, we have $x_1=2$ and $x_2=4$.

The corresponding eigenvalues of $\mathcal{H}$ and $S^z$ are respectively given by
\begin{equation}
    E = \sum_{j=1}^M 2\left( \Delta - \cos(k_j) \right) \,, \qquad S^z = \frac{L}{2} - M \,.
\label{simeigenvalues}
\end{equation}

\subsection{Open XXZ chain}

The Hamiltonian of the $U(1)$-invariant open spin-1/2 XXZ chain is given by \begin{equation}
    \mathcal{H} = -\tfrac{1}{2}\sum_{n=1}^{L-1}
    \left(\sigma^x_n \sigma^x_{n+1} + \sigma^y_n \sigma^y_{n+1} + \Delta \left( \sigma^z_n \sigma^z_{n+1} - \id \right) \right) - \tfrac{1}{2}\left(h\, \sigma^z_1 + h'\, \sigma^z_L \right) + \tfrac{1}{2}(h + h') \id\,,
\label{Hopen}
\end{equation}
where $h$ and $h'$ are boundary magnetic fields. 
The Bethe equations are now given by \cite{Alcaraz:1987uk}
\begin{equation}
\frac{\alpha(k_j)\, \beta(k_j)}{\alpha(-k_j)\, \beta(-k_j)}
= \prod_{l=1; l\ne j}^M \frac{B(-k_j, k_l)}{B(k_j, k_l)} 
\,, \qquad j = 1, \ldots, M \,,
\label{BEopen}
\end{equation}
where 
\begin{equation}
B(k,k') = s(k, k')\, s(k', -k) \,,
\end{equation}  
see \eqref{s(k,k')},
and
\begin{equation}
\alpha(k) = 1 + (h-\Delta)\,e^{-i k} \,, \qquad
\beta(k) = \left[1 + (h'-\Delta)\,e^{-i k} \right] e^{i(L+1) k}\,.  
\end{equation}
The exact simultaneous eigenvectors of $\mathcal{H}$ and $S^z$ again have the form 
$|B^L_M\rangle$ \eqref{BLM} with 
\begin{equation}
    f(w) = \sum_{\sigma\in S_M}
    \sum_{\epsilon_1,\dots,\epsilon_M=\pm 1}
    \varepsilon(\sigma)\,\epsilon_1\dots\epsilon_M\, A(\epsilon_1 k_{\sigma(1)}, \ldots, \epsilon_M k_{\sigma(M)})\, 
    e^{i \sum_{j=1}^M \epsilon_j k_{\sigma(j)} x_j} \,,
\label{fXXZopen}
\end{equation} 
and now
\begin{equation}
A(k_1, \ldots, k_M) = \prod_{j=1}^{M} \beta(-k_j) 
\prod_{1\le j<l\le M} B(-k_j, k_l)\, e^{-i k_l} \,.
\label{Acoeffopen}
\end{equation}
The $x_j \in \{1, \ldots,L\} $ in \eqref{fXXZopen} are again the positions of the 1's in the argument $w$ of $f(w)$.
The corresponding eigenvalues of $\mathcal{H}$ and $S^z$
are again given by \eqref{simeigenvalues}.

\section{Circuit diagrams}\label{sec:examples}

We present here the complete circuit diagrams for examples with $(L,M)=(4,2)$ and $(L,M)=(6,3)$.


\begin{figure}[htb]
      \centering
\includegraphics[width=1\linewidth]{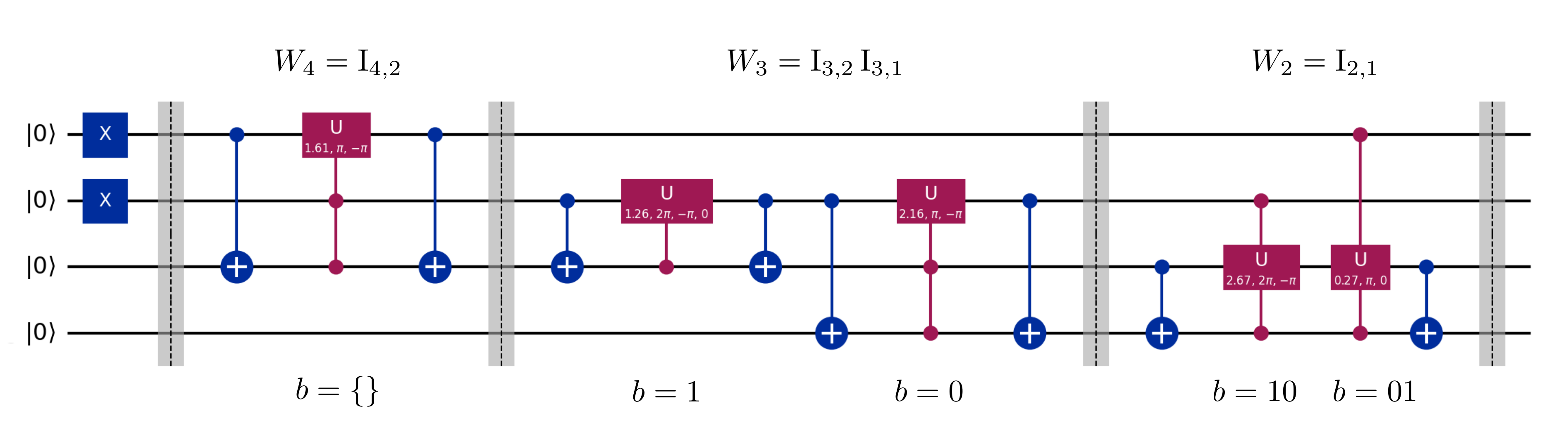}
\caption{The complete circuit diagram for the open XXZ Bethe state 
with $\Delta=0.5,\, h=0.1,\,h'=0.3$, see \eqref{Hopen},
for the case $L=4,\,M=2$, and Bethe roots $\{0.682741,\, 1.38561\}$. Here 
the barriers separate the different $W_m$, see \eqref{ULresult} and 
\eqref{Wmresult}.}
\label{fullcircuit4,2}
\end{figure}
 

\begin{figure}[htb]
      \centering
\includegraphics[width=1\linewidth]{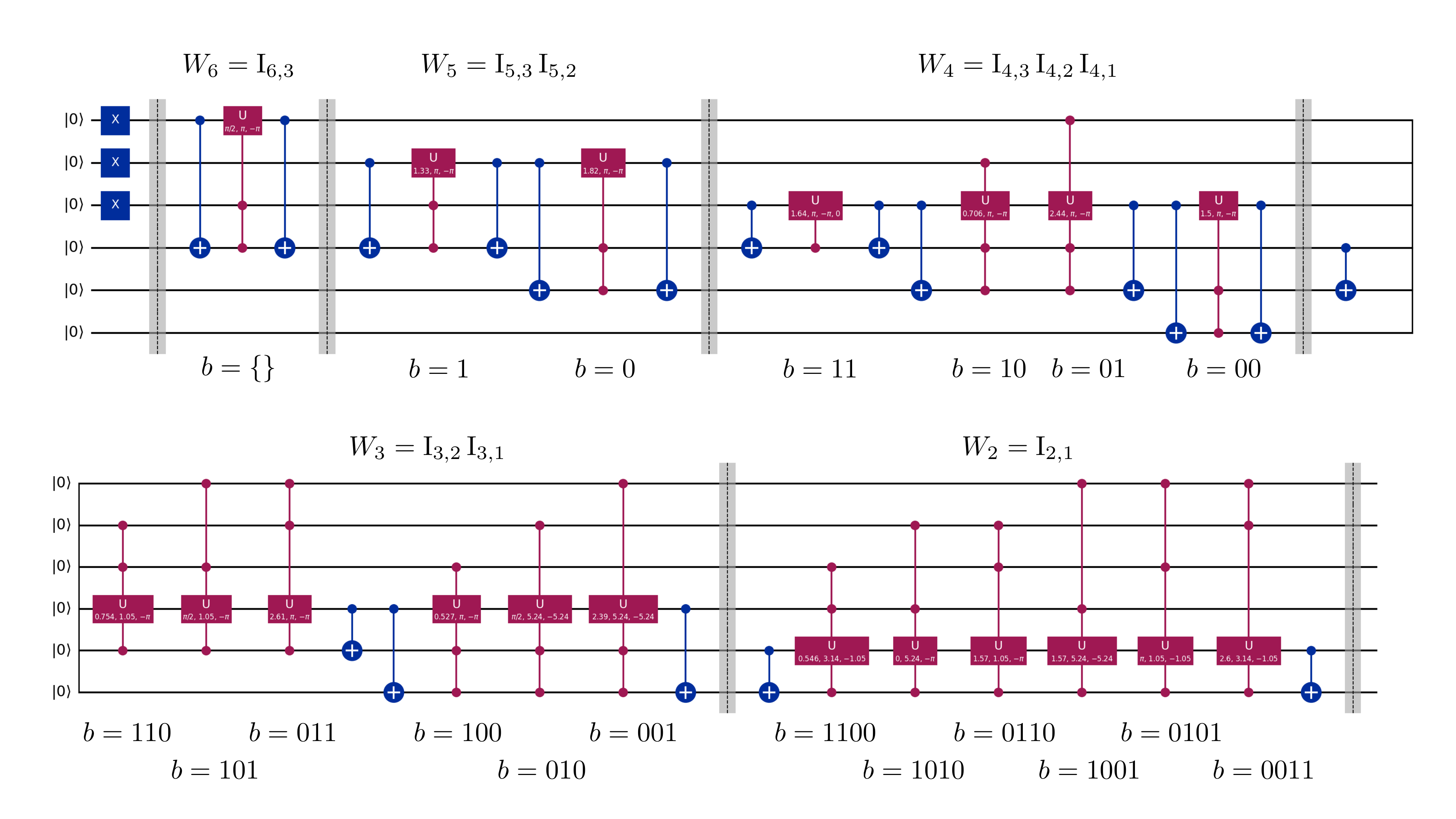}
\caption{The complete circuit diagram for the closed XXZ Bethe state
with $\Delta=1.005$, see \eqref{Hclosed}, for the case $L=6,\,M=3$,
and Bethe roots $\{0.0112138,\,1.04159 -0.7291i,\,1.04159+0.7291i\}$.
Here the barriers separate the different $W_m$, see \eqref{ULresult} and 
\eqref{Wmresult}.}
\label{fullcircuit6,3}
\end{figure}

\clearpage

\end{document}